\documentclass[%
reprint,
 amsmath,amssymb,
 aps,
prl
]{revtex4-1}
\DeclareMathOperator{\sech}{sech}
\DeclareMathAlphabet\EuScript{U}{eus}{m}{n}
\SetMathAlphabet\EuScript{bold}{U}{eus}{b}{n}

\usepackage{graphicx}
\usepackage{dcolumn}
\usepackage{bm}

\begin{document}


\title{Demonstration of a Sensitive Method to Measure Nuclear Spin-Dependent Parity Violation}

\author{Emine Altunta\c{s}}
\email[]{emine.altuntas@yale.edu}
\altaffiliation[Present Address: ] {Joint Quantum Institute, National Institute of Standards and Technology, and University of Maryland, Gaithersburg, MD 20899}
\author{Jeffrey Ammon}
\altaffiliation[Present Address: ] {Lincoln Laboratory, Massachusetts Institute of Technology, Lexington, MA 02420}
\author{Sidney B. Cahn}
\email[]{sidney.cahn@yale.edu}
\affiliation{Department of Physics, P.O.Box 208120, Yale University, New Haven, Connecticut 06520, USA}
\author{David DeMille}
\email[]{david.demille@yale.edu}
\affiliation{Department of Physics, P.O.Box 208120, Yale University, New Haven, Connecticut 06520, USA}

\date{\today}

\begin{abstract}
Nuclear spin-dependent parity violation (NSD-PV) effects in atoms and molecules arise from $Z^0$ boson exchange between electrons and the nucleus, and from the magnetic interaction between electrons and the parity-violating nuclear anapole moment. We demonstrate measurements of NSD-PV that use an enhancement of the effect in diatomic molecules, here using the test system $^{138}$Ba$^{19}$F. Our sensitivity surpasses that of any previous atomic PV measurement. We show that systematic errors can be suppressed to at least the level of the present statistical sensitivity. We measure the matrix element, $W$, of the NSD-PV interaction with total uncertainty $\delta W/(2\pi)<0.7$ Hz, for each of two configurations where $W$ must have different signs. This sensitivity would be sufficient to measure NSD-PV effects of the size anticipated across a wide range of nuclei.  	
\end{abstract}


\maketitle


In atoms and molecules, certain aspects of the electroweak interaction are manifested by nuclear spin dependent parity violation (NSD-PV). NSD-PV primarily arises from two fundamental causes. The first one is the coupling between vector-electron and axial-nucleon neutral currents ($\ensuremath{V_eA_n}$) resulting from $Z^0$ boson exchange. Prior $\ensuremath{V_eA_n}$ measurements at low momentum transfer were made with electron-nucleus scattering; their results are expressed in terms of the dimensionless constants, $C_{2u,d}$, that characterize the $\ensuremath{V_eA_n}$ coupling to up and down quarks~\cite{Slac, Pvdis, Sample, ERLER2013119}.
The present experimental uncertainties in $C_{2u,d}$ are $\gtrsim\!70\%$ of the predicted values~\cite{Sample}. Improved $C_{2u,d}$ measurements would provide a new check of the Standard Model. The second source of NSD-PV is the nuclear anapole moment, which arises from weak interactions within the nucleus~\cite{Zeldovich}. This P-odd magnetic moment couples to the spin magnetic dipole of a penetrating electron~\cite{Flambaum,Flambaum1984}. A nonzero nuclear anapole moment has been measured only once~\cite{Wieman}. Additional measurements of anapole moments may enable determination of parameters describing the strength of purely hadronic PV interactions~\cite{Murray, Haxton}, which have proven difficult to measure by other means~\cite{Gardner,HAXTON2013185}.

Here we demonstrate sensitivity to NSD-PV surpassing any previous atomic PV measurement, based on a novel approach using diatomic molecules~\cite{Labzowsky95, FLAMBAUM1985121,FlambaumSushkov}. Due to their rotational structure, molecules with an unpaired electron spin systematically have small energy splittings between opposite-parity hyperfine states, which can be mixed by NSD-PV. A magnetic field can Zeeman-shift these molecular levels to near degeneracy, which enhances the mixing due to NSD-PV and gives much larger PV signals than with atoms. We measure the strength of the NSD-PV induced mixing using a Stark interference technique~\cite{Nguyen,DeMille}. For many molecules with one valence electron in a $^2\Sigma$ state, measurements can be interpreted in terms of the underlying weak-interaction physics with $\lesssim\!10\%$ accuracy~\cite{Borschevsky, Nayak, Labzowsky95, Isaev, Kozlov}.

Here, we demonstrate our method using the molecule barium monofluoride (BaF), specifically the isotopologue $^{138}$Ba$^{19}$F. NSD-PV effects are nonzero only for nuclei with a nonzero spin, $I$~\cite{DeMille}. Since $I_{\rm Ba} =0$ for $^{138}$Ba, here a NSD-PV signal can arise only from $^{19}$F, where $I_{\rm F}$=1/2. The valence electron wavefunction in BaF has poor overlap with the F nucleus, so the anticipated effect due to $I_F$ is far below our experimental sensitivity~\cite{Private}---i.e., an accurate measurement in this system must be consistent with zero. As such, $^{138}\mathrm{BaF}$ is a powerful system to identify systematic errors. Here we demonstrate control over systematics at a level sufficient for future measurements in many molecular species (including $^{137}$BaF) where NSD-PV effects are nonzero.

The ground electronic state $X^2\Sigma$ of $^{138}\mathrm{BaF}$ is described by the effective Hamiltonian $H = {B_e\bm{N^2}} + {\gamma\bm{N\cdot S}} + {b \bm{I\cdot S}} + c(\bm{I\cdot n})(\bm{S\cdot n})$, where $N$ is the rotational angular momentum, $S=1/2$ is the electron spin, $\gamma$ is the spin-rotation (SR) constant, $b,c$ are hyperfine (HF) constants, and $\bm{n}$ is a unit vector along the internuclear axis ($\hbar=1$ throughout). The field-free eigenstates have energy $E_N \approx B_eN(N+1)$ and parity $P=(-1)^N$, where $B_e$ is the rotational constant. We Zeeman shift sublevels of the $N^P=0^+$ and $1^-$ states to near degeneracy, using a magnetic field $\boldsymbol{\EuScript{B}}=\EuScript{B}\hat{z}$. Zeeman shifts are dominated by the coupling to $\bm{S}$, with approximate Hamiltonian $H_Z\cong-g\mu_B\bm{S}\cdot\boldsymbol{\EuScript{B}}$, where $g\cong-2$ and $\mu_B$ is the Bohr magneton. Since $B_e\gg\gamma,b,c$, the $\mathcal{B}$-field necessary to bridge the rotational energy ${E_1}-{E_0}\approx\!2B_e$ is large enough to strongly decouple $\bm{S}$ from $\bm{I}$ and $\bm{N}$. Thus we write the molecular states in the decoupled basis $| N, m_N \rangle |S, m_s \rangle | I, m_I \rangle$. Level crossings between different pairs of opposite-parity states occur at slightly  different values of $\mathcal{B}$ because of small energy differences from $\gamma$, $b$, and $c$~\cite{DeMille}.

In $^{138}\mathrm{BaF}$, the opposite parity levels $ |0,0\rangle|\frac{1}{2},\frac{1}{2}\rangle|\frac{1}{2},m_I\rangle\equiv| \psi_{\uparrow}^{+} (m_N=0, m_I)\rangle$ and $|1,m_N^{'}\rangle|\frac{1}{2},-\frac{1}{2}\rangle|\frac{1}{2},m_I^{'}\rangle\equiv
| \psi_{\downarrow}^{-} (m_N^{'}, m_I^{'})\rangle$ are degenerate when $\mathcal{B}={\mathcal{B}}_0 \approx\! B_e/\mu_B\sim0.5$ T. The NSD-PV Hamiltonian is a pseudoscalar that mixes levels with opposite parity and the same value of the total angular momentum projection $m\equiv m_S + m_N + m_I$. We use the crossings where $m_I =  m_I^\prime = 1/2$, $m_N^\prime = 1$, and $m$ = 1 (crossing A), and where $m_I = m_I^\prime = -1/2$, $m_N^\prime = 1$, and $m = 0$ (crossing F), in our measurements~\cite{CahnPRL}.


The effective NSD-PV Hamiltonian 
for this system is 
$\label{hamiltonian2}\hat{H}_p^{Eff} =\kappa' W_P C$. Here, $W_P$ is an energy that characterizes the strength of the electron-nucleus overlap, which can be accurately calculated for many species, including BaF~\cite{DeMille, Private}. The dimensionless operator ${C\equiv\left(\bm{n} \times \bm{S}\right)\cdot \bm{I}/{I}}$ gives the angular momentum dependence of $\hat{H}_p^{Eff}$. The dimensionless parameter $\kappa^\prime\approx\kappa_2^\prime + \kappa_a^\prime$ encodes the physics of the weak interactions that lead to NSD-PV; it has contributions primarily from the $V_eA_n$ interaction ($\kappa_2^\prime$) and from the electron-nuclear anapole moment interaction ($\kappa_a^\prime$). We seek to determine $\kappa'$ by measuring the NSD-PV matrix elements $iW(m_N^{'},m_I^{'},m_N,m_I)\equiv\kappa' W_P \tilde{C}$, where $\tilde{C} \equiv \langle \psi_{\downarrow}^{-}(m_N^{'}, m_I^{'}) |C |\psi_{\uparrow}^{+} (m_N, m_I) \rangle$ is determined from standard angular-momentum algebra. ($iW$ is pure imaginary due to time-reversal invariance.) 

\begin{figure}
\includegraphics[width=83mm]{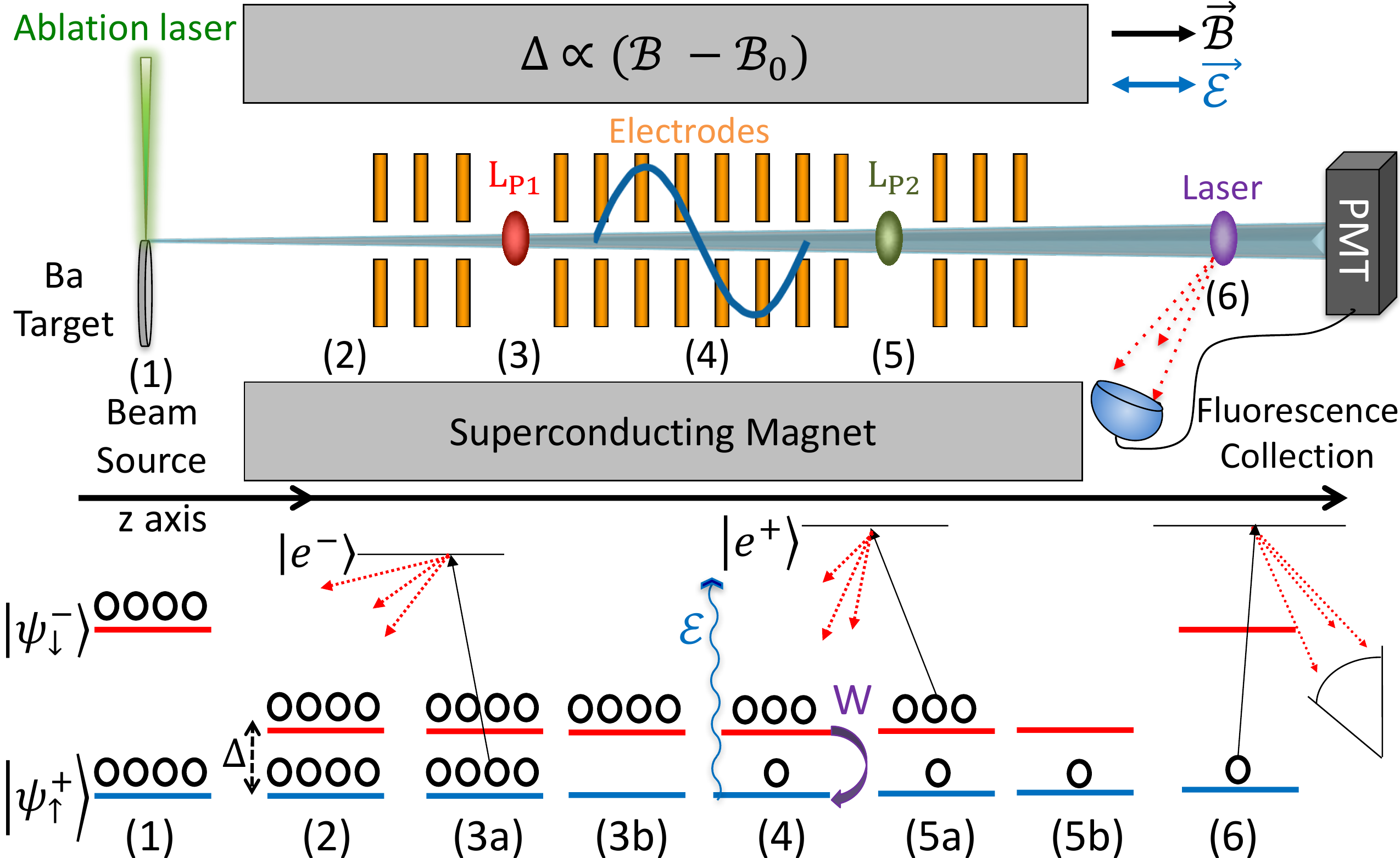}
\caption{(color online). Schematic of the apparatus (top) and evolution of the level populations (bottom). (1) BaF molecules are formed by laser ablation into a pulsed jet; both parity states have equal thermal populations. (2) Molecules enter the magnet, where the states are nearly degenerate. (3) Laser beam $\rm{L}_{P1}$ depletes the $|\psi^+_\uparrow\rangle$ state. (4) A single-cycle sine wave $\mathcal{E}$-field (blue) is applied parallel to $\boldsymbol{\mathcal{B}}$. Stark and NSD-PV interactions mix the opposite parity states and transfer population into $|\psi^+_\uparrow\rangle$. (5) Laser beam $\rm{L}_{P2}$ depletes the $|\psi^-_\downarrow\rangle$ state. (6) Molecules exit the magnet, and population transferred to $|\psi^+_\uparrow\rangle$ is detected. }
\label{fig:ApparatusSchematic}
\end{figure}

The measurement sequence for $W$ proceeds as follows. A beam of $^{138}\mathrm{BaF}$ molecules enters a magnet, where $\EuScript{B}\approx \EuScript{B}_0$ (Fig.~\ref{fig:ApparatusSchematic}). 
A laser beam ($\rm{L}_{P1}$) depletes state $|\psi^+_\uparrow\rangle$, by optically pumping to a short-lived, odd-parity excited state, $|e^-\rangle$, at time $t=0$. After a free evolution time $T_{f1}$, molecules enter a spatially varying electric field, $\boldsymbol{\mathcal{E}} = \mathcal{E}_0 \sin(2\pi z/L)\hat{z}$ for $0<z<L$. Molecules with velocity $\textbf{v}=v\hat{z}$ experience a time-dependent field $\mathcal{E}(t=z/v) = \mathcal{E}_0 \sin(\omega t)$, with $\omega=2\pi v/L\equiv 2\pi/T_e$. Due to the combined Stark and NSD-PV mixing of the levels, population is transferred from the populated state $|\psi_\downarrow^-\rangle$ to the originally emptied state $|\psi_\uparrow^+\rangle$. After the $\mathcal{E}$-field ends, molecules experience a second free evolution time, $T_{f2}$.  Then, another laser beam ($\rm{L}_{P2}$) depletes the odd parity $|\psi^-_\downarrow\rangle$ state by optical pumping via an even-parity excited state, $|e^+\rangle$. For both depletion lasers, selection rules ensure that one parity eigenstate of the unresolved ground state pair is excited, leaving behind a definite parity eigenstate. We refer to $\rm{L}_{P1}$ ($\rm{L}_{P2}$) as the 1$^{\rm st}$ (2$^{\rm nd}$) parity state projection laser. We measure the population of $|\psi_\uparrow^+\rangle$ after the molecules exit the magnet, using laser-induced fluorescence. Although there are conceptual subtleties regarding the evolution of the states as they leave the magnet~\cite{LongPaper, AltuntasThesis}, this description accurately captures the effect of our measurement sequence.

The Hamiltonian for the near-degenerate states, in the basis of parity eigenstates, is~\cite{Flambaum1985} 
\begin{align}
	H_{\pm} =& \left( \begin{array}{c} 0 \\ 
		-iW+d\mathcal{E}(t) \end{array} \right. & \left.
	\begin{array}{c} iW+d\mathcal{E}(t) \\ \Delta\end{array}\right),
	\label{Hplusminus}
\end{align}
where $\Delta$ is the small $\mathcal{B}$-field dependent detuning from exact degeneracy, and $d$ is the dipole matrix element~\cite{DeMille}. The wavefunction is 
\begin{equation}
| \psi\left(t\right)\rangle = c_+\left(t\right)|\psi_{\uparrow}^{+}\rangle + e^{-i\Delta t} c_-\left(t\right)|\psi_{\downarrow}^{-}\rangle \equiv \left(\begin{array}{c} c_+\left(t\right) \\ c_-\left(t\right)\end{array}\right),
\label{Equ:Shro}
\end{equation}
where $c_+\left(0\right) =0$ and $c_-\left(0\right) =1$ due to the optical pumping. For $W\ll d\mathcal{E}_0$, the Schr\"{o}dinger equation at $t=T=T_{f1}+T_e+T_{f2}$ yields
\begin{eqnarray}
c_+\left(T\right) &&= \frac{iW}{\Delta}\left[e^{-i\Delta T}-1\right] \nonumber \\ &&+  \frac{2 d\mathcal{E}_0 \omega}{\omega^2 -\Delta^2}e^{-i\Delta \left(T_e/2+T_{f1}\right)}\sin\left[\frac{\Delta T_e}{2}\right].
\end{eqnarray}
The measured signal ${S= N_0 |c_+\left(T\right)|^2}$, with $N_0$ the initial population of the $|\psi_\uparrow^+\rangle$ state, is given by
\begin{equation}
\begin{split}
S & \simeq  4N_0\left(\frac{ d\mathcal{E}_0 \omega}{\omega^2 -\Delta^2}\right)\left\{\frac{ d\mathcal{E}_0 \omega}{\omega^2 -\Delta^2} \sin^2\left[\frac{\Delta T_e}{2}\right] \right.\\
&\left.+ 2\frac{W}{\Delta} \sin\left[\frac{\Delta T_e}{2}\right] \sin\left[\frac{\Delta}{2}{ T}\right]\cos\left[\frac{\Delta}{2}{\left(T_{f1}-T_{f2}\right)}\right]\right\}.
\end{split}
\end{equation}
With this expression we compute the ``theoretical'' NSD-PV asymmetry, $\mathcal{A}_{thy}$, associated with reversal of $\mathcal{E}_0$: 
\begin{eqnarray}
&&\mathcal{A}_{thy}(\Delta)= \frac{S\left(+\mathcal{E}_0\right)-S\left(-\mathcal{E}_0\right)}{S\left(+\mathcal{E}_0\right)+S\left(-\mathcal{E}_0\right)} = 2\frac{W}{\Delta}\frac{\omega^2 -\Delta^2}{ d\mathcal{E}_0 \omega}\frac{1}{\sin\left[\frac{\Delta}{2}{ T_e}\right]}\nonumber\\
&& \times \sin\left[\frac{\Delta}{2}{ \left(T_e+T_{f1}+T_{f2}\right)}\right] \cos\left[\frac{\Delta}{2}{\left(T_{f1}-T_{f2}\right)}\right].
\label{AsymParity}
\end{eqnarray}
Here, $\Delta$ and $\mathcal{E}_0$ are experimental values we control, the parameters $\omega,T_e, T_{f1}$, and $T_{f2}$ are defined by the geometry of the interaction region (IR), the measured molecular velocity $v$ is $v = 616$ m/s, and $d$ was measured previously~\cite{CahnPRL}. For the idealized case where $T_{f1} = T_{f2} = 0$ and $\Delta \ll \omega$, Eqn.~\ref{AsymParity} reproduces the simpler expression in~\cite{DeMille}.


The $\mathcal{B}$-field is generated by a superconducting (SC) magnet equipped with $5$ SC and $14$ room temperature (RT) gradient coils for adjusting the field homogeneity. $\mathcal{B}$-field homogeneity is critical to obtain a well-defined value of $\Delta$ throughout the IR, and to minimize systematic errors. An array of $32$ magnetometers distributed around the magnet center are used for initial $\mathcal{B}$-field mapping with precision $\delta \mathcal{B} / \mathcal{B} \sim 0.5$ ppm \cite{Murphree2007160}. The Z0 RT shim coil, which provides a uniform field, is employed to tune $\Delta$. An additional set of home-built shim coils (the \textquotedblleft mini-shims") creates local variations in $\mathcal{B}(z)$, to shape it for maximum homogeneity and/or to generate gradients for systematic error tests. 

The IR enables control over $\mathcal{E}$-fields and delivery of laser light to the molecular beam inside the magnet. The IR has 32 cylindrical electrodes, all with inner diameter 3.18 cm: 2 long tubes (used as endcaps), 28 rings ($\sim\!6$ mm long), and 2 extra-wide rings ($\sim\!17$ mm long to mount prisms that reflect laser light). For PV data, we apply voltages to these electrodes as required to generate the desired sinusoidal $\mathcal{E}(z)$-field.

We also apply localized $\mathcal{E}$-field pulses to the IR electrodes, to measure $\mathcal{E}$ and $\mathcal{B}$-field inhomogeneities and to study possible systematics. A spatially narrow $\mathcal{E}$-field pulse, centered at position $z_k$, is created by applying uniform voltages $V_0$ for all electrodes at $z<z_k$, and $V_0 + \delta V$ at $z>z_k$. This unipolar field pulse is approximated by $\mathcal{E}^u(t;t_k) ={\mathcal{E}}_{0}^{u} \sech([t-t_k]/\sigma_u)$, where $\mathcal{E}_{0}^{u}=0.42~\delta \rm V$/cm, $\sigma_u=0.76$ cm$/v$ and $t-t_k=(z-z_k)/v$. In general, for any weak electric field $\mathcal{E}(t)$ (and ignoring the effect of $W$), the signal $S\propto |\tilde{\mathcal{E}}(\Delta)|^2$, where $\tilde{\mathcal{E}}(\Delta)$ is the Fourier transform of ${\mathcal{E}}\left(t\right)$~\cite{CahnPRL}. For the unipolar pulse, the signal is $S^u(\Delta) \propto \sech^2\left(\pi \Delta\sigma_u/2\right)$. 
 

\begin{figure}
\includegraphics[width=83mm]{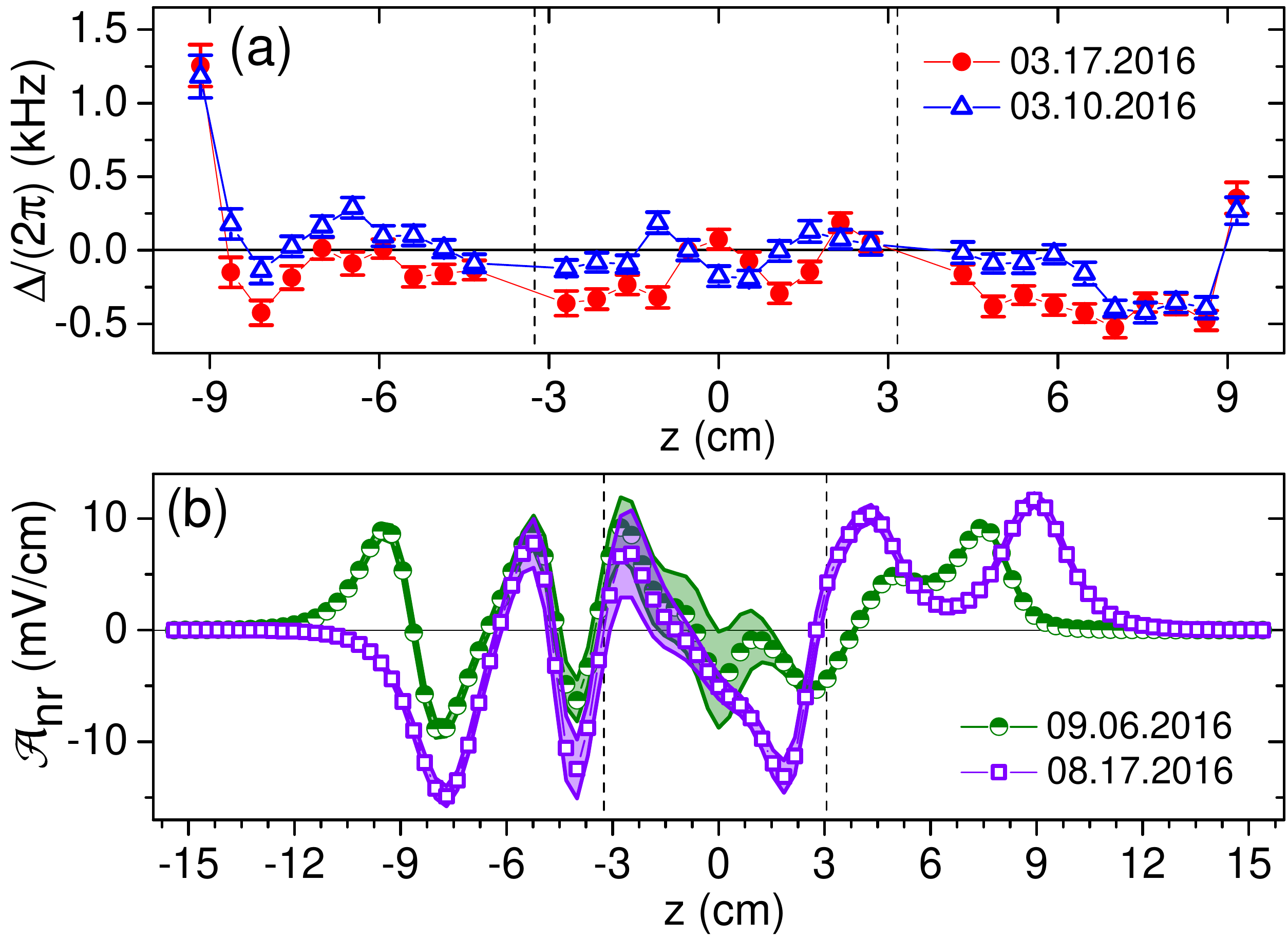}
\caption{(color online) Typical $\mathcal{B}$- and $\mathcal{E}$-fields after shimming. Vertical dashed lines indicate the region of interest, between the $L_{\rm P1,2}$ beam locations. (a) Magnetic field variations $\delta \mathcal{B}$ vs. $z$, in terms of $\Delta$ via the relation $\Delta\approx 2\mu_B\mathcal{B}$, for measurements taken 1 week apart. (b) Shimmed ambient non-reversing $\mathcal{E}$-field, $\mathcal{A}_{nr}$, vs. $z$ from measurements taken many days apart. Shaded areas indicate $\pm 1 \sigma$ statistical uncertainty range.}
\label{fig:AnrBfieldPeformance}
\end{figure}

To measure the $\mathcal{B}$-field, we apply unipolar $\mathcal{E}$-field pulses centered at a series of discrete locations, $z_k$, and at each $z_k$ find the detuning required for exact level crossing, ${\Delta}_k = 2 {\mu}_B ({\mathcal{B}}_k -{\mathcal{B}}_0)$, from the peak of $S^u(\Delta)$. Here, ${\mathcal{B}}_k$ is the actual $\mathcal{B}$-field at $z_k$, and ${\mathcal{B}}_0$ is the field required for ${\Delta}=0$. We then shim the $\mathcal{B}$-field to minimize the variance within the set of $\Delta_k$ values. We routinely achieve $\delta\mathcal{B}/\mathcal{B} \lesssim 2\times 10^{-8}$ (r.m.s.) after shimming (Fig.~\ref{fig:AnrBfieldPeformance}(a)), corresponding to $\delta\Delta \lesssim 2\pi\times200$ Hz.


To measure stray, non-reversing $\mathcal{E}$-fields, ${\mathcal{E}}_{nr}$, we apply a larger, reversible unipolar field pulse, $\pm{\mathcal{E}}^u(t; t_k)$. The total field is ${\mathcal{E}}\left(t\right)_{\pm}={\mathcal{E}}_{nr}(t)\pm{\mathcal{E}}^u(t; t_k)$, yielding the signal $S_\pm(\Delta)\propto|\tilde{\cal{E}}_{\pm}(\Delta)|^2$. The difference signal, $S_\delta(\Delta)\equiv S_+(\Delta)-S_-(\Delta)$, arises from interference terms proportional to $\mathcal{E}_{0}^u$ and $\mathcal{E}_{nr}$. By applying the reversible pulses at different locations $z_k$, we determine $\tilde{\cal{E}}_{nr}$; from its inverse Fourier transform, we then find $\mathcal{E}_{nr}(z)$. To shim away non-reversing fields, we generate $-\mathcal{E}_{nr}$ with the IR electrodes. We refer to the ambient non-reversing $\mathcal{E}$-field, after shimming, as $\mathcal{A}_{nr}(z)$, to distinguish it from non-reversing fields under other conditions. Typically $\langle \mathcal{A}_{nr}\rangle_{\rm r.m.s.} < 6$ mV/cm (Fig.~\ref{fig:AnrBfieldPeformance}(b)).


Our strategy for identifying systematic errors is as follows. We first shim imperfections as described above, and set upper bounds on their residual nonzero values. Next, we deliberately amplify a possible experimental imperfection by a factor $A$ (defined as the ratio of deliberate to maximum ambient imperfection) and observe its coupling to \textit{other} ambient imperfections by measuring the resulting induced offset in $W$. If we find an offset bounded by $|\Delta W|$, we infer that the systematic error in $W$ under ambient conditions is no larger than $|\Delta W|/A$.

The deliberately amplified imperfections included non-reversing $\mathcal{E}$-fields, $\mathcal{B}$-field inhomogeneities, and offsets in laser detunings. We employed two different shapes of deliberately-applied $\mathcal{E}_{nr}$ fields: the previously defined unipolar pulse, $\mathcal{E}^u(t;t_k)$, and a bipolar pulse, generated by applying equal and opposite unipolar pulses at adjacent electrode gaps. For each shape, we performed measurements at several different IR locations. For $\mathcal{B}$-field inhomogeneities, we performed measurements using both the RT shim coils (for large-scale gradients) and the mini-shim coils (for localized $\mathcal{B}$-field pulses). In all measurements, offsets in $W$ with the amplified imperfections present were within $2.1\sigma$ of the null result. 

Only two combinations of imperfections were found to give definite shifts in $W$: (1) a linear $\mathcal{B}$-field gradient together with a unipolar $\mathcal{E}_{nr}$, and (2) a detuning offset of the $\rm{L}_{P2}$ laser ($\delta\nu_{L2}$), together with an $\mathcal{E}_{nr}$ near the $\rm{L}_{P2}$ beam. In the first case, we found an analytic solution to the Schr{\" o}dinger Eqn. that explains the effect. Simultaneous application of a Gaussian $\mathcal{E}_{nr}$, $\mathcal{E}_{nr}^g(t)=\mathcal{E}_1{e^{-\left({t}/{\sigma_g}\right)}}^2$, and a linear $\mathcal{B}$-field gradient, $\partial\mathcal{B}/\partial t \equiv\gamma/(2\mu_B)$, results in an asymmetry $\mathcal{A}_{gsl}$ that mimics the NSD-PV effect: $\mathcal{A}_{gsl}(\Delta)=B_{gsl}\Delta/
\left(\Delta^2+x^2\right)$. Here, $x=\frac{\mathcal{E}_{1}}{\mathcal{E}_{0}}\frac{\sigma_g}{2}\frac{\omega^2}{\sqrt{\pi}}$, $B_{gsl}=\frac{2}{\sqrt{\pi}}\frac{\mathcal{E}_{1}}{\mathcal{E}_{0}}{\gamma\frac{\sigma_g}{2}\left( \frac{\pi^2}{2} -3 - \frac{\sigma_g^2 \omega^2}{4}\right)}$, and we assumed $\Delta , d\mathcal{E}_0 \ll \omega$, $\gamma \ll \omega^2$, $\gamma \ll \sigma_g^{-2}$, and $c_+ \ll 1$. We tested this analytical model by deliberately applying known values of $\mathcal{E}_{nr}^g(t)$ and $\partial{\mathcal{B}}/\partial t$, and found that
the ratio of experimental and predicted values of the asymmetry amplitude, $B_{gsl}$, is $ 1.26 \pm 0.08$. This calibration accuracy of ${\sim\!25}\%$ is sufficient for our current null measurement result with $^{138}$BaF. 

\begin{table}
	\centering
	\begin{tabular}{ p{4.2cm} p{1.5cm}  p{2.3cm}  }
		\hline
		\\[-0.9em]
		Parameter &Shift $\frac{\Delta W}{2\pi}$ & Uncertainty $\frac{\delta W}{2\pi}$ \\
		\\[-0.9em]
		\hline	
		\\[-0.8em]	
		Bipolar $\mathcal{E}_{nr}$ Pulses & &0.12  	\\
		\\[-0.8em]
		Unipolar $\mathcal{E}_{nr}$ Pulses & &0.16  \\
		\\[-0.8em]
		$\mathcal{B}$-Field Inhomogeneities &  &0.24 \\	
		\\[-0.8em]
		Linear $\mathcal{B}$-Field Gradient and & -0.01 & 0.02\\
		Unipolar $\mathcal{E}_{nr}$ at center &  & \\
		\\[-0.8em]
		Detuning offset in $\rm L_{P2}$ and & -0.04  &0.21  \\
		$\mathcal{E}_{nr}$ at same position & & \\
		\hline
		\hline
		\\[-0.6em]
		Total Systematic & -0.05 &0.38   \\
		\hline
		\hline 	 
	\end{tabular}
	\caption{Systematic error budget (all values in Hz). Uncertainties are added in quadrature. }		
	\label{Table:SysErr}
\end{table}

The second combination of imperfections leads to a systematic offset $\Delta W \propto \mathcal{E}_{nr}\cdot\delta\nu_{L2}$. By measuring the proportionality
constant in this relation, and then bounding the maximum laser detuning and the maximum $\mathcal{E}_{nr}$ value during normal operation, we set a limit on the maximum associated systematic error in $W$. Table~\ref{Table:SysErr} presents our final  systematic error budget. 

\begin{figure}
	\includegraphics[width=83mm]{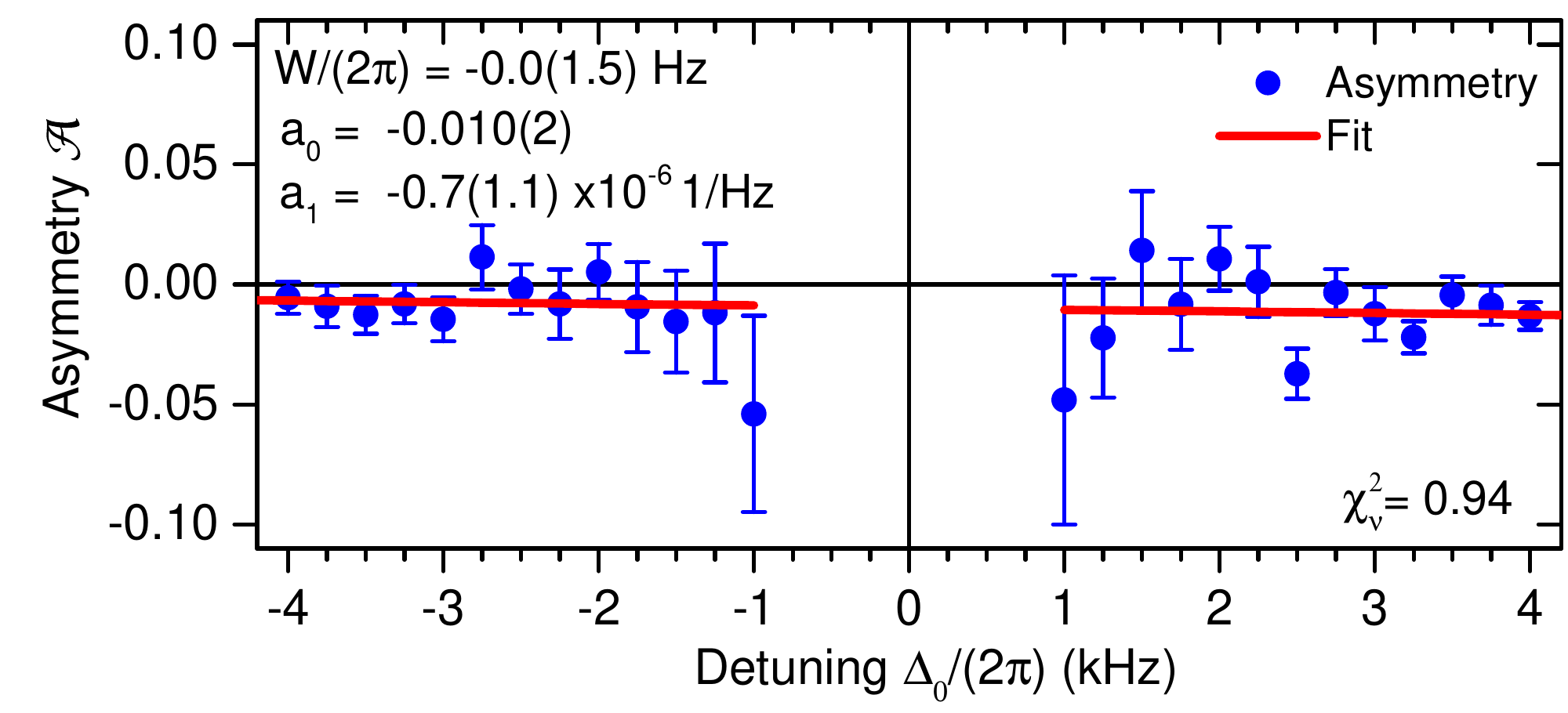}
	\caption{(color online) Sample PV asymmetry data (points) and fit (line). This shows $\approx 4.4$ hours of data.}
	\label{fig:AsymmetryRun3842}	
\end{figure}

Fig.~\ref{fig:AsymmetryRun3842} shows asymmetry data from a typical NSD-PV measurement run. We extract $W$ by fitting the asymmetry to $\mathcal{A}(\Delta)= W_{\rm{fit}} [\mathcal{A}_{thy}(\Delta)/W] +a_0+a_1\Delta$, with $\mathcal{A}_{thy}(\Delta)$ defined in Eqn.~\ref{AsymParity} and
$T_e=2\pi/\omega=87~\mu$s, $T_{f1}=7.4~\mu$s, $T_{f2}=8.9~\mu$s, $\mathcal{E}_0=1$ V/cm, and $d_{A(F)}/(2\pi)=3360~(3530)$ Hz/(V/cm) for the A(F) crossing~\cite{CahnPRL}. The free parameters in the fit are $W_{\rm{fit}}$ plus the auxiliary coefficients $a_0$ and $a_1$, which were found necessary through extensive numerical simulations taking into account likely imperfections in the fields. The offset term $a_0$ is induced by $\mathcal{E}_{nr}$ fields alone, and
is found in both simulations and data to be uncorrelated with systematic errors in $W$. The auxiliary parameter $a_1$ \textit{is} strongly associated with systematic errors in $W$; a nonzero $a_1$ serves as a preliminary diagnostic of such errors.

\begin{figure}	
	\includegraphics[scale=0.98]{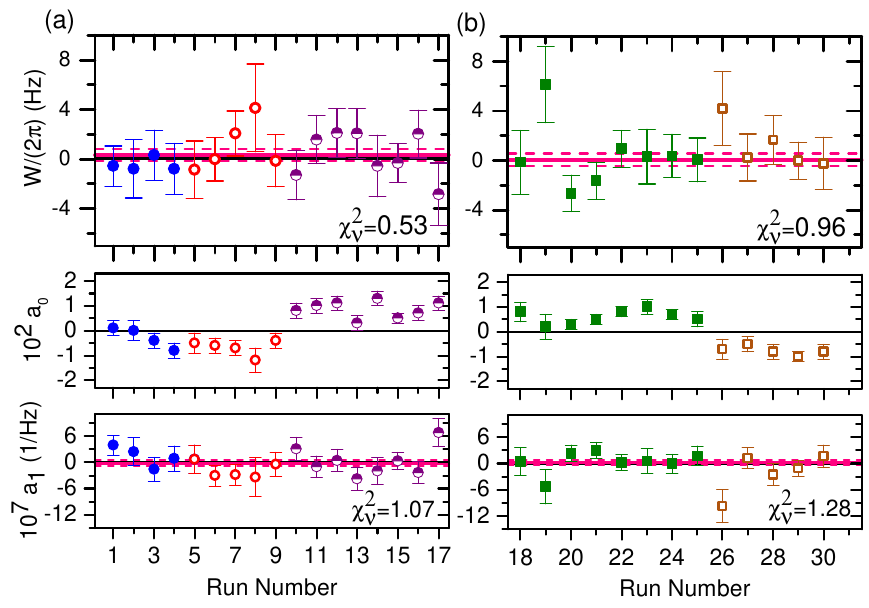}
	\caption{(color online) Summary of NSD-PV data. Data taken with the same $\mathcal{E}_{nr}$ shim voltages are denoted with the same symbol. Error bars represent 1$\sigma$ statistical uncertainties. (a)/(b) NSD-PV data runs at crossing A/F. Top: Results for $W$. Weighted averages and uncertainties denoted by red solid and dashed lines. Middle: $a_0$ results. Note the consistency within a given $\mathcal{E}_{nr}$ shim voltage set, but differences between sets. Bottom: $a_1$ results, with weighted averages $a_{1}^{(A)}=-2(6)$ and $a_{1}^{(F)}=3(6)\times10^{-8}$/Hz. }
	\label{fig:AllPVonlyA0BothCrossings}
\end{figure}

We performed NSD-PV measurements with several different $\mathcal{E}_{nr}$ shim voltages to check for any dependence on a specific form of the ambient field $\mathcal{A}_{nr}$. As seen in Fig.~\ref{fig:AllPVonlyA0BothCrossings}, measured $W$ values for different $\mathcal{A}_{nr}$ fields are consistent, as expected. Moreover, the average $a_1$ value is consistent with zero in all data, as expected in the absence of significant systematic errors. As a further test for systematics, we compared results from both level crossings, A and F. The value of the factor $\tilde{C}$ differs between crossings, while the factor $W_{\rm mol} \equiv \kappa^\prime W_P$ is the same at all crossings. Here, $\tilde{C}_A=-0.41i$ and $\tilde{C}_F=0.39i$, so $iW= W_{\rm mol}\tilde{C}$ must change sign between crossings. By contrast, the dipole matrix elements $d_{A(F)}$ have the same sign and similar magnitude at both crossings. Thus, systematics due to a common $\mathcal{E}$-field imperfection would give a systematic shift of the same sign at both crossings, unlike a true NSD-PV signal.

Our final results at each crossing are $W^{(A)}/(2\pi)=0.29(53)(41)$ Hz and $W^{(F)}/(2\pi)=0.00(55)(41)$ Hz, yielding  $W_{\text{mol}}/(2\pi)=-0.39(0.95)(1.02)$ Hz, where the first error is statistical, and the second systematic. Systematic errors are evaluated for the entire set of measurements (not for each crossing individually); thus, the total systematic uncertainty is the simple average of the individual contributions. Since the statistical errors are not correlated, we average them in quadrature.
Combining statistical and systematic errors in quadrature, our final result is $W_{\text{mol}}/(2\pi)\equiv\kappa' W_P/(2\pi)=-0.39(1.40)~\text{Hz}$. A more detailed description of our experiment and its results is given in~\cite{LongPaper}. 



In $^{138}$BaF, our result can be interpreted as a measurement of $\kappa^{\prime}(^{19}\rm F)$. The calculated overlap of the valence electron with the $^{19}$F nucleus yields $W_{P}(\rm F)/(2\pi)=0.05$ Hz~\cite{Private}, so that $\kappa^{\prime}(^{19}\rm F)=-8(28)$. A simple nuclear shell-model prediction yields $\kappa^\prime_{\rm thy}(^{19}\rm F) \approx -0.08$~\cite{Flambaum, Flambaum1984, DeMille, Kopfermann}. The consistency of our measurement with this near-zero predicted value demonstrates the absence of systematic errors outside our uncertainty.

Determining $\kappa^{\prime}(^{19}\rm F)$ was not the primary goal of this study. It is more useful to compare our sensitivity to that of previous atomic PV experiments, and to projections for future molecular NSD-PV measurements. With $\lesssim\!90$ hours of data at each crossing, our statistical uncertainty for $W^{(A)}$ or $W^{(F)}$ is $\delta W/(2\pi)< 0.6$ Hz. The previous most sensitive atomic PV experiment, using Dy, had $\delta W/(2\pi) =2.9$ Hz with $\sim\!30$ hours of data~\cite{Nguyen}. We next aim to measure NSD-PV in $^{137}$BaF, where $W_{P}(\rm Ba)/(2\pi) = 160(15)$ Hz~\cite{Borschevsky, DeMille, Nayak,Labzowsky95,Isaev,Titov}. The crude expectation is ${\kappa_{\rm thy}^\prime{(^{137}\rm Ba)}\approx0.07}$, with roughly equal contributions expected from $\kappa'_a$ and $\kappa'_2$~\cite{DeMille}. With the same uncertainties as here, the projected uncertainty would be $\delta\kappa'{(^{137}\rm Ba)}=0.009$, sufficient for a $\sim\!10\%$ measurement. Our projected uncertainty in $^{137}\rm BaF$ would represent a factor of ${\sim\!7}$ improvement relative to the atomic measurement of NSD-PV in $^{133}$Cs, where $\delta\kappa'\!\approx0.06$~\cite{Wieman}. Our technique is sufficiently general and already sensitive enough to enable measurements across a broad range of diatomic molecules~\cite{DeMille}. This gives the promise of determining purely hadronic PV interaction strengths~\cite{Haxton01}. Future measurements with our technique also may be useful for constraining the strength of PV interactions mediated by lighter analogues of the $Z^0$ boson~\cite{Dzuba2017}.

\begin{acknowledgments}
This work was supported by the National Science Foundation under grant no. PHY-1404162.  We thank M.G.~Kozlov, D.~Murphree, D.A.~Rahmlow, E.~Kirilov, Y.V.~Gurevich, and R.~Paolino for contributions to earlier stages of this work.
\end{acknowledgments}

\bibliography{PRL_138BaF_v4Jan18}

\end{document}